\documentclass[journal]{IEEEtran}

\usepackage{graphicx} \usepackage{bm} \usepackage{amsmath}
\usepackage{cite}
\usepackage{amssymb} 

\usepackage{amsmath,amsthm,amssymb,mathrsfs}
\usepackage{bm}
\usepackage{subfigure}

\newcommand{\D}{{\rm d}}

\ifCLASSINFOpdf
\else
\fi
\begin{document}

\title{Theoretical Study of Spin-Torque Oscillator with Perpendicularly Magnetized Free Layer}
\author{Tomohiro~Taniguchi,
        Hiroko~Arai,
        Hitoshi~Kubota,
        and~Hiroshi~Imamura$^{*}$
        \\
        Spintronics Research Center, AIST, 
        Tsukuba, Ibaraki 305-8568, Japan
\thanks{${}^{*}$Corresponding author. Email address: h-imamura@aist.go.jp}}

\maketitle

\begin{abstract}
The magnetization dynamics of spin torque oscillator (STO) 
consisting of a perpendicularly magnetized free layer 
and an in-plane magnetized pinned layer was studied 
by solving the Landau-Lifshitz-Gilbert equation. 
We derived the analytical formula of the relation between 
the current and the oscillation frequency of the STO 
by analyzing the energy balance between 
the work done by the spin torque and 
the energy dissipation due to the damping. 
We also found that the field-like torque breaks the energy balance, 
and change the oscillation frequency. 
\end{abstract}


\IEEEpeerreviewmaketitle


\section{Introduction}
\label{sec:Introduction}

\IEEEPARstart{S}{pin} 
torque oscillator (STO) has attracted much attention due to its potential uses for 
a microwave generator and 
a recording head of a high density hard disk drive. 
The self-oscillation of the STO was first discovered 
in an in-plane magnetized 
giant-magnetoresistive (GMR) system [1]. 
After that, the self-oscillation of the STO has been observed not only in GMR systems [2]-[6] 
but also in magnetic tunnel junctions (MTJs) [7]-[11]. 
The different types of STO have been proposed recently; 
for example, a point-contact geometry with a confined magnetic domain wall [12]-[14] 
which enables us to control the frequency 
from a few GHz to a hundred GHz. 


Recently, Kubota \textit{et al.} experimentally developed the MgO-based MTJ 
consisting of a perpendicularly magnetized free layer and 
an in-plane magnetized pinned layer [15],[16]. 
They also studied the self-oscillation of this type of MTJ, 
and observed a large power ($\sim 0.5$ $\mu$W) with a narrow linewidth ($\sim 50$ MHz) [17]. 
These results are great advances in realizing the STO device. 
However, the relation between the current and the oscillation frequency still remains unclear. 
Since a precise control of the oscillation frequency of the STO by the current is 
necessary for the application, 
it is important to clarify 
the relation between the current and the oscillation frequency. 



In this paper, 
we derived the theoretical formula of 
the relation between the current and the oscillation frequency of the STO 
consisting of the perpendicularly magnetized free layer and the in-plane magnetized pinned layer. 
The derivation is based on the analysis of 
the energy balance between 
the work done by the spin torque and the energy dissipation due to the damping. 
We found that the oscillation frequency monotonically decreases with increasing the current 
by keeping the magnetization in one hemisphere of the free layer. 
The validity of the analytical solution was confirmed by numerical simulations. 
We also found that the field-like torque breaks the energy balance, 
and change the oscillation frequency. 
The shift direction of the frequency, 
high or low, is determined by the sign of the field-like torque. 


This paper is organized as follows. 
In Sec. II, 
the current dependence of the oscillation frequency is derived 
by solving the Landau-Lifshitz-Gilbert (LLG) equation. 
In Sec. III, 
the effect of the field-like torque 
on the oscillation behaviour is investigated. 
Section IV is devoted to the conclusions. 


\begin{figure}
  \centerline{\includegraphics[width=0.7\columnwidth]{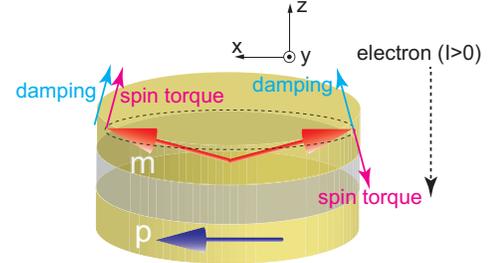}}
  \caption{
           Schematic view of the system. 
           The directions of the spin torque and the damping 
           during the precession around the $z$-axis are indicated. 
  \vspace{-3.5ex}}
  \label{fig:fig1}
\end{figure}




\section{LLG study of spin torque oscillation}
\label{sec:LLG study of spin torque oscillation}

The system we consider is schematically shown in Fig. \ref{fig:fig1}. 
We denote the unit vectors pointing in the directions of 
the magnetization of the free and the pinned layers 
as $\mathbf{m}=(\sin\theta\cos\varphi,\sin\theta\sin\varphi,\cos\theta)$ and $\mathbf{p}$, respectively. 
The $x$-axis is parallel to $\mathbf{p}$ 
while the $z$-axis is normal to the film plane. 
The variable $\theta$ of $\mathbf{m}$ is the tilted angle from the $z$-axis 
while $\varphi$ is the rotation angle from the $x$-axis. 
The current $I$ flows along the $z$-axis, 
where the positive current corresponds to the electron flow 
from the free layer to the pinned layer. 


We assume that the magnetization dynamics is well described by 
the following LLG equation: 
\begin{equation}
  \frac{{\rm d}\mathbf{m}}{{\rm d}t}
  =
  -\gamma
  \mathbf{m}
  \times
  \mathbf{H}
  -
  \gamma
  H_{\rm s}
  \mathbf{m}
  \times
  \left(
    \mathbf{p}
    \times
    \mathbf{m}
  \right)
  +
  \alpha
  \mathbf{m}
  \times
  \frac{\D\mathbf{m}}{\D t}.
  \label{eq:LLG}
\end{equation}
The gyromagnetic ration and the Gilbert damping constant are denoted as 
$\gamma$ and $\alpha$, respectively. 
The magnetic field is defined by 
$\mathbf{H}=-\partial E/\partial (M \mathbf{m})$, 
where the energy density $E$ is 
\begin{equation}
  E
  =
  -MH_{\rm appl}
  \cos\theta
  -
  \frac{M(H_{\rm K}-4\pi M)}{2}
  \cos^{2}\theta.
  \label{eq:energy}
\end{equation}
Here, $M$, $H_{\rm appl}$, and $H_{\rm K}$ are 
the saturation magnetization, 
the applied field along the $z$-axis, 
and the crystalline anisotropy field along the $z$-axis, respectively. 
Because we are interested in the perpendicularly magnetized system, 
the crystalline anisotropy field, 
$H_{\rm K}$, should be larger than the demagnetization field, $4\pi M$. 
Since the LLG equation conserves the norm of the magnetization, 
the magnetization dynamics can be described by a trajectory on an unit sphere. 
The equilibrium states of the free layer correspond to $\mathbf{m}=\pm \mathbf{e}_{z}$. 
In following, 
the initial state is taken to be the north pole, 
i.e., $\mathbf{m}=\mathbf{e}_{z}$. 
It should be noted that 
a plane normal to the $z$-axis, in which $\theta$ is constant, corresponds to 
the constant energy surface. 


The spin torque strength, $H_{\rm s}$ in Eq. (\ref{eq:LLG}), is [18]-[20]
\begin{equation}
  H_{\rm s}
  =
  \frac{\hbar \eta I}{2e (1 + \lambda m_{x})MSd}, 
  \label{eq:H_s}
\end{equation}
where $S$ and $d$ are the cross section area and the thickness of the free layer. 
Two dimensionless parameters, $\eta$ and $\lambda$ ($-1 < \lambda < 1$), determine 
the magnitude of the spin polarization 
and the angle dependence of the spin torque, respectively. 
Although the relation among $\eta$, $\lambda$, and the material parameters depends 
on the theoretical models [20]-[22], 
the form of Eq. (\ref{eq:H_s}) is applicable 
to both GMR system and MTJs. 
In particular, the angle dependence of the spin torque 
characterized by $\lambda$ is a key 
to induce the self-oscillation in this system. 


\begin{figure}
  \centerline{\includegraphics[width=1.0\columnwidth]{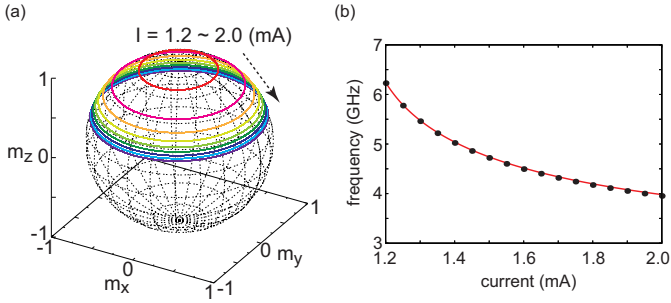}}
  \caption{
           (a) The trajectories of the steady state precession of the magnetization in the free layer 
               with various currents. 
           (b) The dots represent the dependence of the oscillation frequency 
               obtained by numerically solving the LLG equation. 
               The solid line is obtained by Eqs. (\ref{eq:I_theta}) and (\ref{eq:frequency}). 
  \vspace{-3.5ex}}
  \label{fig:fig2}
\end{figure}


Figure \ref{fig:fig2} (a) shows the steady state precession of the magnetization in the free layer 
obtained by numerically solving Eq. (\ref{eq:LLG}). 
The values of the parameters are 
$M=1313$ emu/c.c., 
$H_{\rm K}=17.9$ kOe, 
$H_{\rm appl}=1.0$ kOe, 
$S=\pi \times 50 \times 50$ nm${}^{2}$, 
$d=2.0$ nm,
$\gamma=17.32$ MHz/Oe, 
$\alpha=0.005$, 
$\eta=0.33$, 
and $\lambda=0.38$, 
respectively [17]. 
The self-oscillation was observed for the current $I \ge 1.2$ mA. 
Although the spin torque breaks the axial symmetry of the free layer along the $z$-axis, 
the magnetization precesses around the $z$-axis 
with an almost constant tilted angle. 
The tilted angle from the $z$-axis increases 
with increasing the current; 
however, the magnetization stays in the north semisphere ($\theta < \pi/2$). 
The dots in Fig. \ref{fig:fig2} (b) show the dependence of the oscillation frequency on the current. 
As shown, the oscillation frequency monotonically decreases 
with increasing the current magnitude. 



Let us analytically derive the relation between the current and the oscillation frequency. 
Since the self-oscillation occurs 
due to the energy supply into the free layer by the spin torque, 
the energy balance between the spin torque and the damping should be investigated. 
By using the LLG equation, 
the time derivative of the energy density $E$ is given by 
${\rm d}E/{\rm d}t=\mathcal{W}_{\rm s}+\mathcal{W}_{\alpha}$, 
where
the work done by spin torque, $\mathcal{W}_{\rm s}$, and 
the energy dissipation due to the damping, $\mathcal{W}_{\alpha}$, are respectively 
given by
\begin{equation}
  \mathcal{W}_{\rm s}
  =
  \frac{\gamma MH_{\rm s}}{1+\alpha^{2}}
  \left[
    \mathbf{p}
    \cdot
    \mathbf{H}
    -
    \left(
      \mathbf{m}
      \cdot
      \mathbf{p}
    \right)
    \left(
      \mathbf{m}
      \cdot
      \mathbf{H}
    \right)
    -
    \alpha
    \mathbf{p}
    \cdot
    \left(
      \mathbf{m}
      \times
      \mathbf{H}
    \right)
  \right],
  \label{eq:W_s}
\end{equation}
\begin{equation}
  \mathcal{W}_{\alpha}
  =
  -\frac{\alpha\gamma M}{1+\alpha^{2}}
  \left[
    \mathbf{H}^{2}
    -
    \left(
      \mathbf{m}
      \cdot
      \mathbf{H}
    \right)^{2}
  \right].
  \label{eq:W_alpha}
\end{equation}


By assuming a steady precession around the $z$-axis 
with a constant tilted angle $\theta$, 
the time averages of $\mathcal{W}_{\rm s}$ and $\mathcal{W}_{\alpha}$ 
over one precession period are, respectively, given by 
\begin{equation}
\begin{split}
  \overline{\mathcal{W}}_{\rm s}
  =&
  \frac{\gamma M}{1+\alpha^{2}}
  \frac{\hbar \eta I}{2e \lambda MSd}
  \left(
    \frac{1}{\sqrt{1-\lambda^{2}\sin^{2}\theta}}
    -
    1
  \right)
\\
  &\times
  \left[
    H_{\rm appl}
    +
    \left(
      H_{\rm K}
      -
      4\pi M
    \right)
    \cos\theta
  \right]
  \cos\theta,
  \label{eq:W_s_ave}
\end{split}
\end{equation}
\begin{equation}
  \overline{\mathcal{W}}_{\alpha}
  =
  -\frac{\alpha \gamma M}{1+\alpha^{2}}
  \left[
    H_{\rm appl}
    +
    \left(
      H_{\rm K}
      -
      4\pi M
    \right)
    \cos\theta
  \right]^{2}
  \sin^{2}\theta.
  \label{eq:W_alpha_ave}
\end{equation}
The magnetization can move from the initial state to a point at which $\overline{{\rm d}E/{\rm d}t}=0$. 
Then, the current at which a steady precession with the angle $\theta$ can be achieved is given by 
\begin{equation}
\begin{split}
  I(\theta)
  =&
  \frac{2 \alpha e \lambda MSd}{\hbar \eta \cos\theta}
  \left(
    \frac{1}{\sqrt{1-\lambda^{2}\sin^{2}\theta}}
    -
    1
  \right)^{-1}
\\
  &\times
  \left[
    H_{\rm appl}
    +
    \left(
      H_{\rm K}
      -
      4\pi M
    \right)
    \cos\theta
  \right]
  \sin^{2}\theta.
  \label{eq:I_theta}
\end{split}
\end{equation}
The corresponding oscillation frequency is given by 
\begin{equation}
  f(\theta)
  =
  \frac{\gamma}{2\pi}
  \left[
    H_{\rm appl}
    +
    \left(
      H_{\rm K}
      -
      4\pi M
    \right)
    \cos\theta
  \right].
  \label{eq:frequency}
\end{equation}
Equations (\ref{eq:I_theta}) and (\ref{eq:frequency}) are the main results 
in this section. 
The solid line in Fig. \ref{fig:fig2} (b) shows the current dependence of the oscillation frequency 
obtained by Eqs. (\ref{eq:I_theta}) and (\ref{eq:frequency}), 
where the good agreement with the numerical results confirms the validity of the analytical solution. 
The critical current for the self-oscillation, 
$I_{\rm c}=\lim_{\theta \to 0}I(\theta)$, is given by 
\begin{equation}
  I_{\rm c}
  =
  \frac{4 \alpha e MSd}{\hbar \eta \lambda}
  \left(
    H_{\rm appl}
    +
    H_{\rm K}
    -
    4\pi M 
  \right). 
  \label{eq:I_c}
\end{equation}
The value of $I_{\rm c}$ estimated by using the above parameters is $1.2$ mA, 
showing a good agreement with the numerical simulation shown in Fig. \ref{fig:fig2} (a). 
The sign of $I_{\rm c}$ depends on that of $\lambda$, 
and the self-oscillation occurs only for the positive (negative) current 
for the positive (negative) $\lambda$. 
This is because a finite energy is supplied to the free layer for $\lambda \neq 0$, 
i.e., $\overline{\mathcal{W}}_{\rm s} >0$. 
In the case of $\lambda=0$, 
the average of the work done by the spin torque is zero, 
and thus, 
the self-oscillation does not occur. 


It should be noted that $I(\theta) \to \infty$ in the limit of $\theta \to \pi/2$. 
This means 
the magnetization cannot cross over the $xy$-plane, 
and stays in the north hemisphere ($\theta < \pi/2$). 
The reason is as follows. 
The average of the work done by spin torque becomes zero 
in the $xy$-plane ($\theta=\pi/2$) 
because the direction of the spin torque is parallel to the constant energy surface. 
On the other hand, the energy dissipation due to the damping is finite 
in the presence of the applied field [21]. 
Then, $\overline{{\rm d}E/{\rm d}t}(\theta=\pi/2)=-\alpha\gamma M H_{\rm appl}^{2}/(1+\alpha^{2}) < 0$, 
which means the damping moves the magnetization to 
the north pole. 
Thus, the magnetization cannot cross over the $xy$-plane. 
The controllable range of the oscillation frequency by the current is 
$f(\theta=0)-f(\theta=\pi/2)=\gamma (H_{\rm K}-4\pi M)/(2\pi)$, 
which is independent of the magnitude of the applied field. 


Since the spin torque breaks the axial symmetry of the free layer along the $z$-axis, 
the assumption that the tilted angle is constant 
used above is, in a precise sense, not valid, 
and the $z$-component of the magnetization oscillates around a certain value. 
Then, the magnetization can reach the $xy$-plane and stops its dynamics 
when a large current is applied. 
However, the value of such current is more than 15 mA for our parameter values, 
which is much larger than 
the maximum of the experimentally available current. 
Thus, the above formulas work well in the experimentally conventional current region. 


Contrary to the system considered here, 
the oscillation behaviour of 
an MTJ with an in-plane magnetized free layer and a perpendicularly magnetized pinned layer 
has been widely investigated [23]-[26]. 
The differences of the two systems are as follows. 
First, the oscillation frequency decreases with increasing the current in our system 
while it increases in the latter system. 
Second, the oscillation frequency in our system in the large current limit 
becomes independent of 
the $z$-component of the magnetization 
while it is dominated by $m_{z}=\cos\theta$ in the latter system. 
The reasons are as follows. 
In our system, 
by increasing the current, 
the magnetization moves away from the $z$-axis 
due to which the effect of the anisotropy field on the oscillation frequency decreases, 
and the frequency tends to $\gamma H_{\rm appl}/(2\pi)$, 
which is independent of the anisotropy. 
On the other hand, 
in the latter system, 
the magnetization moves to the out-of-plane direction, 
due to which the oscillation frequency is strongly affected by the anisotropy (demagnetization field). 


The macrospin model developed above reproduces the experimental results 
with the free layer of 2nm thick [17], 
for example the current-frequency relation, quantitatively. 
Although only the zero-temperature dynamics is considered in this paper, 
the macrospin LLG simulation at a finite temperature also reproduces other properties, 
such as the power spectrum and its linewidth, well. 
However, when the free layer thickness further decreases, 
an inhomogeneous magnetization 
due to the roughness at the MgO interfaces may affects the magnetization dynamics: 
for example, 
a broadening of the linewidth. 



\section{Effect of field-like torque}
\label{sec:Effct of field-like torque}


The field-like torque arises from the spin transfer 
from the conduction electrons to the local magnetizations, 
as is the spin torque. 
When the momentum average of the transverse spin 
of the conduction electrons relaxes in the free layer very fast, 
only the spin torque acts on the free layer [19]. 
On the other hand, when the cancellation of the transverse spin is insufficient, 
the field-like torque appears. 
The field-like torque added to the right hand side of Eq. (\ref{eq:LLG}) is 
\begin{equation}
  \mathbf{T}_{\rm FLT}
  =
  -\beta 
  \gamma
  H_{\rm s} 
  \mathbf{m}
  \times
  \mathbf{p},
  \label{eq:FLT}
\end{equation}
where the dimensionless parameter $\beta$ characterizes 
the ratio between the magnitudes of the spin torque and the field-like torque. 
The value and the sign of $\beta$ depend on the system parameters 
such as the band structure, the thickness, the impurity density, 
and/or the surface roughness [22],[27]-[29]. 
The magnitude of the field-like torque in MTJ is much larger than that in GMR system [30],[31] 
because the band selection during the tunneling leads to 
an insufficient cancellation of the transverse spin by the momentum average.


\begin{figure}
  \centerline{\includegraphics[width=0.9\columnwidth]{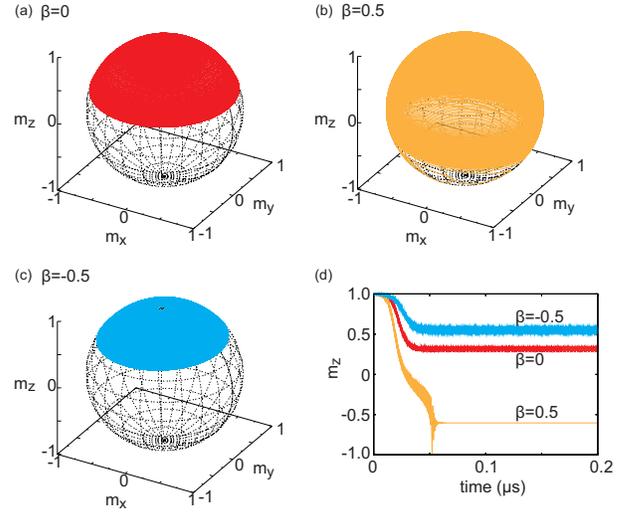}}
  \caption{
           The magnetization dynamics from $t=0$ 
           with (a) $\beta=0$, (b) $\beta=0.5$, and $\beta=-0.5$. 
           The current magnitude is $2.0$ mA. 
           (d) The time evolutions of $m_{z}$ for various $\beta$. 
  \vspace{-3.5ex}}
  \label{fig:fig3}
\end{figure}


It should be noted that 
the effective energy density, 
\begin{equation}
  E_{\rm eff}
  =
  E
  -
  \beta M 
  \frac{\hbar \eta I}{2e \lambda MSd}
  \log
  \left(
    1
    +
    \lambda
    m_{x}
  \right),
  \label{eq:energy_eff}
\end{equation}
satisfying $-\gamma \mathbf{m}\times\mathbf{H}+\mathbf{T}_{\rm FLT}=-\gamma \mathbf{m} \times [-\partial E_{\rm eff}/(M \mathbf{m})]$, 
can be introduce to describe the field-like torque. 
The time derivative of the effective energy, $E_{\rm eff}$, can be obtained 
by replacing the magnetic field, $\mathbf{H}$, in Eqs. (\ref{eq:W_s}) and (\ref{eq:W_alpha}) 
with the effective field $-\partial E_{\rm eff}/\partial (M \mathbf{m})=\mathbf{H}+\beta H_{\rm s} \mathbf{p}$. 
Then, the average of ${\rm d}E_{\rm eff}/{\rm d}t$ over one precession period around the $z$-axis consists of 
Eq. (\ref{eq:W_s_ave}), (\ref{eq:W_alpha_ave}), 
and the following two terms:
\begin{equation}
\begin{split}
  \overline{\mathcal{W}}_{\rm s}^{\prime}
  = &
  \frac{\beta\gamma M}{1 + \alpha^{2}}
  \left(
    \frac{\hbar \eta I}{2e \lambda MSd}
  \right)^{2}
  \left[
    \frac{1 + \lambda^{2}\cos 2 \theta}{(1 - \lambda^{2}\sin^{2}\theta)^{3/2}}
    -
    1
  \right],
  \label{eq:W_s_FLT_ave}
\end{split}
\end{equation}
\begin{equation}
\begin{split}
  \overline{\mathcal{W}}_{\alpha}^{\prime}
  = &
  -\frac{\alpha \gamma M}{1 + \alpha^{2}}
  \left(
    \frac{\beta \hbar \eta I}{2e \lambda MSd}
  \right)^{2}
  \left[
    \frac{1 + \lambda^{2}\cos 2 \theta}{(1 - \lambda^{2}\sin^{2}\theta)^{3/2}}
    -
    1
  \right]
\\
  &-
  \frac{2 \alpha \beta \gamma M}{1+\alpha^{2}}
  \frac{\hbar \eta I}{2e \lambda MSd}
  \left(
    \frac{1}{\sqrt{1-\lambda^{2}\sin^{2}\theta}}
    -
    1
  \right)
\\
  &\times
  \left[
    H_{\rm appl}
    +
    \left(
      H_{\rm K}
      -
      4\pi M
    \right)
    \cos\theta
  \right]
  \cos\theta.
  \label{eq:W_alpha_FLT_ave}
\end{split}
\end{equation}
The constant energy surface of $E_{\rm eff}$ shifts from the $xy$-plane 
due to a finite $|\beta| (\simeq 1)$, 
leading to an inaccuracy of the calculation of the time average 
with the constant tilted angle assumption. 
Thus, Eqs. (\ref{eq:W_s_FLT_ave}) and (\ref{eq:W_alpha_FLT_ave}) are quantitatively valid 
for only $|\beta| \ll 1$. 
However, predictions from Eqs. (\ref{eq:W_s_FLT_ave}) and (\ref{eq:W_alpha_FLT_ave}) 
qualitatively show good agreements with the numerical simulations, 
as shown below. 


For positive $\beta$, 
$\overline{\mathcal{W}}_{\rm s}^{\prime}$ is also positive, 
and is finite at $\theta=\pi/2$. 
Thus, $\overline{{\rm d}E_{\rm eff}/{\rm d}t}(\theta=\pi/2)$ can be positive 
for a sufficiently large current.
This means, the magnetization can cross over the $xy$-plane, 
and move to the south semisphere ($\theta>\pi/2$). 
On the other hand, 
for negative $\beta$, 
$\overline{\mathcal{W}}_{\rm s}^{\prime}$ is also negative. 
Thus, the energy supply by the spin torque is suppressed 
compared to the case of $\beta=0$. 
Then, a relatively large current is required 
to induce the self-oscillation with a certain oscillation frequency. 
Also, the magnetization cannot cross over the $xy$-plane. 



\begin{figure}
  \centerline{\includegraphics[width=0.9\columnwidth]{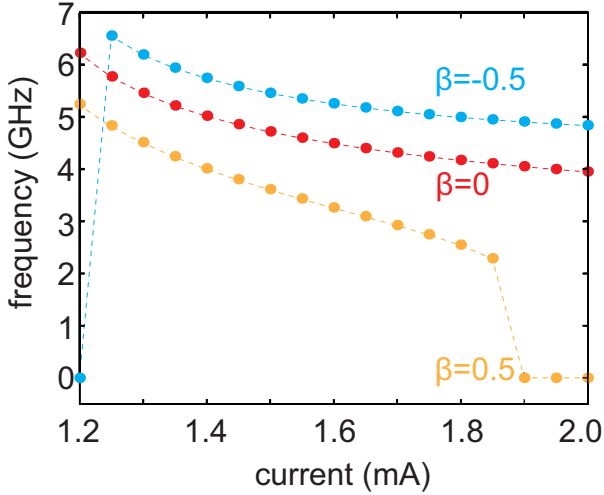}}
  \caption{
           The dependences of the oscillation frequency on the current 
           for $\beta=0$ (red), $\beta=0.5$ (orange), and $\beta=-0.5$ (blue), respectively. 
  \vspace{-3.5ex}}
  \label{fig:fig4}
\end{figure}


We confirmed these expectations by the numerical simulations. 
Figures \ref{fig:fig3} (a), (b) and (c) show 
the trajectories of the magnetization dynamics with $\beta=0$, $0.5$, and $-0.5$ respectively, 
while the time evolutions of $m_{z}$ are shown in Fig. \ref{fig:fig3} (d). 
The current value is 2.0 mA. 
The current dependences of the oscillation frequency are summarized 
in Fig. \ref{fig:fig4}. 


In the case of $\beta=0.5>0$, 
the oscillation frequency is low compared to that for $\beta=0$ 
because the energy supply by the spin torque is enhanced by the field-like torque, 
and thus, the magnetization can largely move from the north pole. 
Above $I=1.9$ mA, 
the magnetization moves to the south hemisphere ($\theta > \pi/2$), 
and stops near $\theta \simeq \cos^{-1}[-H_{\rm appl}/(H_{\rm K}-4\pi M)]$ 
in the south hemisphere, 
which corresponds to the zero frequency in Fig. \ref{fig:fig4}. 


On the other hand, 
in the case of $\beta=-0.5<0$, 
the magnetization stays near the north pole compared to the case of $\beta=0$, 
because the energy supply by the spin torque is suppressed by the field-like torque. 
The zero frequency in Fig. \ref{fig:fig4} indicates 
the increase of the critical current of the self-oscillation. 
Compared to the case of $\beta=0$, 
the oscillation frequency shifts to the high frequency region 
because the magnetization stays near the north pole. 





\section{Conclusions}
\label{sec:Conclusions}

In conclusion, 
we derived the theoretical formula of 
the relation between the current and the oscillation frequency of STO 
consisting of the perpendicularly magnetized free layer and the in-plane magnetized pinned layer. 
The derivation is based on the analysis of 
the energy balance between 
the work done by the spin torque and the energy dissipation due to the damping. 
The validity of the analytical solution was confirmed by 
numerical simulation. 
We also found that the field-like torque breaks the energy balance, 
and changes the oscillation frequency. 
The shift direction of the frequency, high or low, 
depends on the sign of the field-like torque ($\beta$). 


\section*{Acknowledgment}

The authors would like to acknowledge 
T. Yorozu, H. Maehara, A. Emura, M. Konoto, A. Fukushima, S. Yuasa, K. Ando, 
S. Okamoto, N. Kikuchi, O. Kitakami, T. Shimatsu, 
K. Kudo, H. Suto, T. Nagasawa, R. Sato, and K. Mizushima. 


\ifCLASSOPTIONcaptionsoff
  \newpage
\fi


\begin{thebibliography}{20}
\bibitem{IEEEhowto:comment1}
In the absence of the applied field, 
$\overline{{\rm d}E/{\rm d}t}(\theta=\pi/2)=0$. 
In this case, $I(\theta)$ given by Eq. (\ref{eq:I_theta}) satisfies 
$I_{\rm c}=I(\theta=0)>I(\theta=\pi/2)$. 
These results mean that, above the critical current, 
the magnetization immediately moves from the initial state to the $xy$-plane, 
and stop its motion. 
Thus, the self-oscillation does not occur.
\end{thebibliography}


\end{document}